\documentclass[
reprint,
superscriptaddress,
showpacs,preprintnumbers,
nofootinbib,
amsmath,amssymb,
aps,
prd,
floatfix,
]{revtex4-1}

\usepackage{bm}
\usepackage{graphicx}
\usepackage{amssymb}
\usepackage{color}
\usepackage{natbib}
\usepackage{subfigure}
\usepackage{verbatim}
\usepackage{dcolumn}
\usepackage[amssymb]{SIunits}
\usepackage{multirow}
\usepackage{tabularx}
\usepackage{booktabs}
\usepackage{float}
\usepackage[normalem]{ulem}

\newcommand{\be}{\begin{equation}}
\newcommand{\ee}{\end{equation}}
\newcommand{\ba}{\begin{eqnarray}}
\newcommand{\ea}{\end{eqnarray}}

\newcommand{\LCDM}{$\mathrm{\Lambda}$CDM}
\newcommand{\geff}{$\textit{G}_{\rm eff}$}

\begin{document}

\title{$\Lambda$CDM or self-interacting neutrinos? - how CMB data can tell the two models apart}

\author{Minsu~Park}
\email{minsup@princeton.edu}
\affiliation{Department of Physics, Princeton University, Princeton, New Jersey 08544 USA}
\affiliation{Department of Astrophysical Sciences, Princeton University, Princeton, New Jersey 08544 USA}

\author{Christina D.~Kreisch}
\email{ckreisch@astro.princeton.edu}
\affiliation{Department of Astrophysical Sciences, Princeton University, Princeton, New Jersey 08544 USA}

\author{Jo~Dunkley}
\affiliation{Department of Physics, Princeton University, Princeton, New Jersey 08544 USA}
\affiliation{Department of Astrophysical Sciences, Princeton University, Princeton, New Jersey 08544 USA}

\author{Boryana~Hadzhiyska}
\affiliation{Department of Physics, Harvard University, Cambridge, Massachusetts 02138, USA}

\author{Francis-Yan~Cyr-Racine}
\affiliation{Department of Physics, Harvard University, Cambridge, Massachusetts 02138, USA}
\affiliation{Department of Physics and Astronomy, University of New Mexico,1919 Lomas Blvd NE, Albuquerque, New Mexico 87131, USA}
\date{\today}

\begin{abstract}
  Of the many proposed extensions to the \LCDM~paradigm, a model in which neutrinos self-interact until close to the epoch of matter-radiation equality has been shown to provide a good fit to current cosmic microwave background (CMB) data, while at the same time alleviating tensions with late-time measurements of the expansion rate and matter fluctuation amplitude. Interestingly, CMB fits to this model either pick out a specific large value of the neutrino interaction strength, or are consistent with the extremely weak neutrino interaction found in \LCDM, resulting in a bimodal posterior distribution for the neutrino self-interaction cross section. In this paper, we explore why current cosmological data select this particular large neutrino self-interaction strength, and by consequence, disfavor intermediate values of the self-interaction cross section. We show how it is the $\ell \gtrsim 1000$ CMB temperature anisotropies, most recently measured by the {\it Planck} satellite, that produce this bimodality. We also establish that smaller scale temperature data, and improved polarization data measuring the temperature-polarization cross-correlation, will best constrain the neutrino self-interaction strength. We forecast that the upcoming Simons Observatory should be capable of distinguishing between the models.
\end{abstract}
\maketitle

\section{introduction}
 Within the Standard Model of particle physics, neutrinos remain elusive. While universally present, their weak interactions with other particles make them difficult to study directly. Neutrino oscillation experiments have shown that neutrinos have mass \cite{sno, superk}, but the Standard Model does not account for the mechanism that generates this mass \cite{mass1, mass2, mass3, mass4}. This presents the neutrino sector as an intriguing source of new physics.

In the Standard Model we assume neutrinos interact only through the electroweak interaction and decouple from the cosmic plasma at a temperature of 1.5 MeV \cite{1.5}. Once decoupled, the neutrinos freely streamed through the early universe, interacting only through gravity. The free-streaming of these gravitationally coupled neutrinos imposes a shear stress on the matter as it streams past, damping acoustic oscillations in the photon-baryon plasma  and boosting CDM fluctuations at horizon entry \cite{shear}.  Recent observations of the Cosmic Microwave Background (CMB), most recently by the {\it Planck} satellite \cite{planckparams:2018}, have put bounds on neutrino parameters, including the effective number of neutrino species ($N_{\rm eff}=2.99\pm0.17$) and the sum of the species' masses ($\mathrm{\sum m_\nu<0.24~eV}$ at 95\% confidence, or $\mathrm{\sum m_\nu<0.12~eV}$ combined with baryon acoustic oscillation data). This is approaching the lower mass limit for the inverted neutrino hierarchy, $\sum m_\nu \ge 0.1$~eV,  from neutrino oscillation experiments \cite{sno, superk}. Cosmological data can now put competitive constraints on neutrino physics.

New neutrino interactions have become a topic of increasing interest due their impact on cosmological observables via altering neutrino free-streaming during the radiation dominated era (see e.g. Refs.~\cite{BialynickaBirula:1964zz,Bardin:1970wq,Gelmini:1980re,Chikashige:1980qk,Barger:1981vd,Raffelt:1987ah,Kolb:1987qy,Konoplich:1988mj,Berkov:1988sd,Belotsky:2001fb, Hannestad:2004qu,Chacko:2004cz,Hannestad:2005ex,Sawyer:2006ju,Mangano:2006mp,Friedland:2007vv,Hooper:2007jr,Serra:2009uu,Aarssen:2012fx,Jeong:2013eza,Laha:2013xua,Archidiacono:2014nda,Ng:2014pca,Cherry:2014xra,Archidiacono:2015oma,Cherry:2016jol,Archidiacono:2016kkh,Dvali:2016uhn,Capozzi:2017auw,Brust:2017nmv,Forastieri:2017bkq,neutrinophilicDM,Lorenz:2018fzb,Choi:2018gho}). Past studies \cite{FCY, kreisch, lachlan, 2014paper, 2017paper, inflation} have explored the viability of stronger neutrino self-scattering, using a Yukawa interaction model parameterized by an interaction strength, \geff. Here the rate of scattering, $\mathrm{\Gamma_\nu}$, scales as $\mathrm{\Gamma_\nu \propto} \textit{G}_{\rm eff}^2\mathrm{T_\nu^5}$ where $\mathrm{T_\nu}$ is the temperature of the cosmic neutrino background \cite{FCY,lachlan, kreisch, 2014paper, 2017paper}. Increasing \geff\ strengthens the neutrino-neutrino coupling in the early Universe. Thus, increasing \geff\ ultimately delays neutrino free-streaming to epochs of lower energies and lower redshifts. 

\begin{figure*}[ht!]
\includegraphics[width=\linewidth]{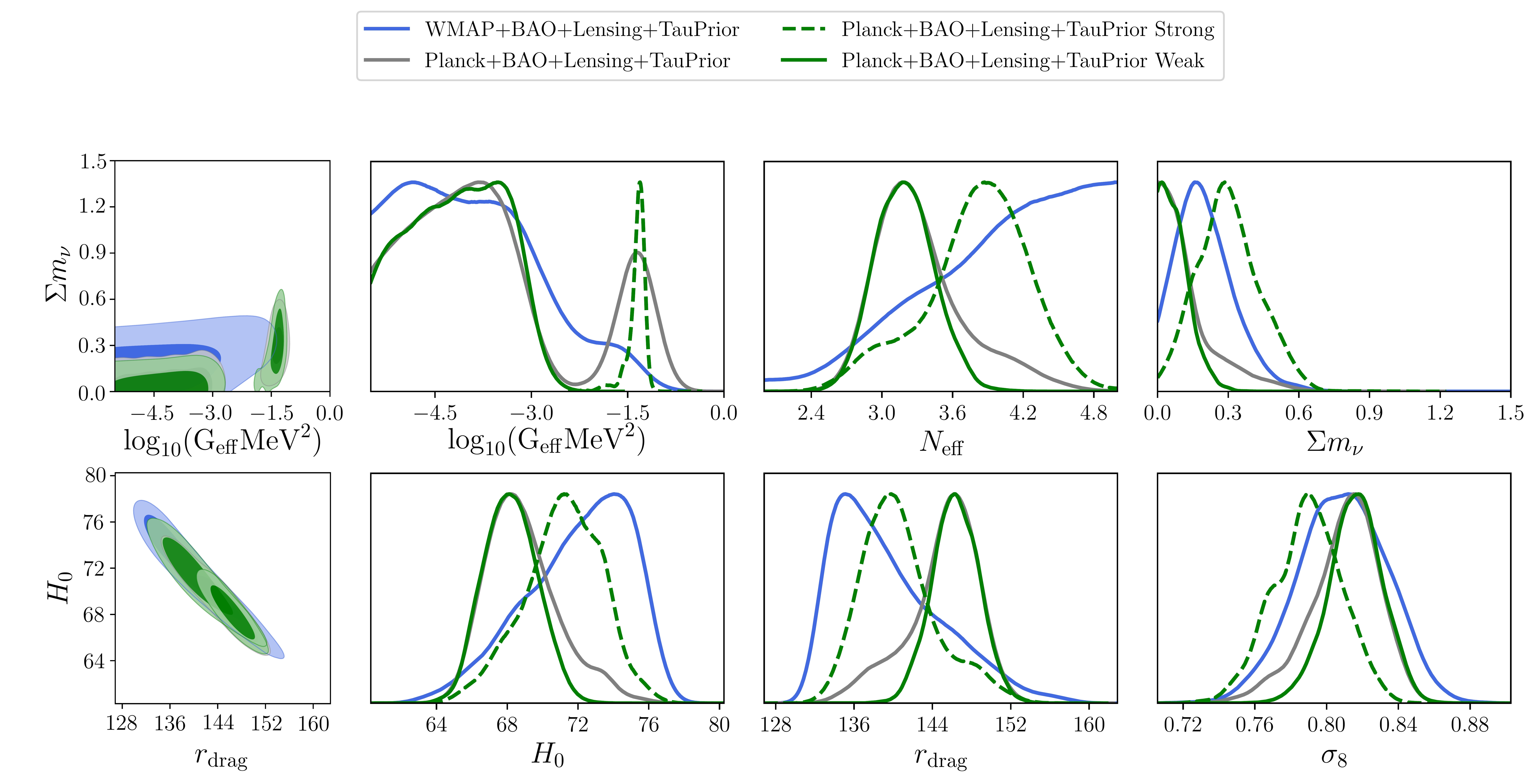}
\caption{Probability distributions for parameters from a nine-parameter model (\LCDM\ plus neutrino self-interaction strength \geff, effective neutrino number, and neutrino mass), using the {\it WMAP} and {\it Planck} CMB data combined with BAO and {\it Planck} lensing data. The parameters derived using {\it Planck} are consistent with previous results \citep{kreisch} and show the clear bimodality in the neutrino self-interaction strength. The `strong' and `weak' distributions show the marginalized posteriors when considering each of the bimodal islands separately. For the unseparated distribution, the strong mode has a lower marginalized posterior relative to the weak mode. The distribution using just {\it WMAP} data is not bimodal.}
\label{fig:planckparams}
\end{figure*}

A delay in the onset of neutrino free-streaming affects the amplitude and phases of the CMB power spectrum (see \cite{kreisch, fullref} for more details.). A model with non-zero value of \geff\ has been shown to fit current cosmological data and produce a bimodal posterior probability for the interaction rate: a `weak' mode with low self-scattering strength \geff, essentially indistinguishable from no self-scattering (\LCDM), and a `strong' self-interacting mode with \geff\ of order $10^{-2}$~MeV$^{-2}$, where the neutrinos decouple at neutrino temperatures as low as 25 eV \cite{FCY, kreisch, lachlan}. The strong mode is particularly interesting as it has a larger predicted Hubble constant \cite{riess} than the usual \LCDM\ model, and a lower predicted amplitude of structure \cite{HSC}, which are preferred by other astronomical datasets.

In this paper we further explore which aspects of current CMB data produce the degeneracy between the two models. We identify the part of the {\it Planck} data responsible for producing the bimodality, which was not present with the {\it WMAP} data, and show how the data exclude models with moderate self-interaction. We then assess how upcoming CMB data might distinguish between the two models. This extends similar investigations in \cite{lachlan}.

\section{Methods}

We use MCMC methods to map out the posterior distribution for a 9-parameter cosmological model: 6 parameters are the usual \LCDM\ parameters (baryon density, cold dark matter density, angular peak position, spectral index and amplitude, and optical depth) and we also vary the effective number of neutrino species, $N_{\rm eff}$, the sum of neutrino masses, $\sum m_\nu$, and the interaction strength \geff. We impose linear priors on all parameters, except \geff\ which takes a logarithmic prior. This prior choice is further discussed in Section \ref{subsec:param}. We use the CosmoMC sampling code \cite{cosmomc} with Multinest \cite{multinest}, which is well-suited to multimodal posteriors.  We use the same modified Boltzmann code, CAMB, as in \cite{kreisch}, and implement the same modifications in the CLASS code as a cross-check.

The datasets used are {\it Planck} 2015 temperature and lensing likelihood using the Plik-lite code \cite{plancklike:2015, plancklens:2015}, combined with current BAO data \cite{BAO1, BAO2, BAO3}, and a gaussian prior on the optical depth of $\tau=0.058\pm0.012$ from {\it Planck}. We also examine the effect of replacing just the {\it Planck} TT data with the {\it WMAP} 9-year TT and TE data \cite{bennett/etal:2013}, using the same BAO data and optical depth prior. Additionally, we generate simulated TT, TE and EE spectra representative of the upcoming Simons Observatory (SO), with co-added white noise levels of 5${\rm \mu}$K-amin over 40\% of the sky, a 1.4$^\prime$ beam and maximum multipoles of $\ell=3000$ in temperature and $\ell=5000$ in polarization \cite{so}\footnote{In this study we do not include the non-white noise and residual foregrounds considered in \cite{so}.}.  We describe the input models for these simulations in Sec.~\ref{subsec:forecast}. 

\section{Results}

\subsection{Parameter distributions with current data}
\label{subsec:param}
In Figure \ref{fig:planckparams} we show a set of the posterior distributions for the sampled and derived parameters for the {\it Planck} data compared to the {\it WMAP} data. Both data were accompanied by the same BAO data and $\tau$ prior. For {\it Planck} we find results consistent with \cite{kreisch,lachlan}, with a bimodal distribution for \geff. One mode is consistent with \LCDM, and the other `strong' mode has non-zero interactions. We identify the preferred parameters for each mode by plotting them separately in Figure \ref{fig:planckparams}, in addition to the joint distribution. The weak mode has $\rm{log}(\textit{G}_{\rm{eff}}{\rm MeV}^2) < -3.04$, $\sum m_\nu < 0.2$eV, and $N_{\rm eff}= 3.19^{+0.51}_{-0.48}$ at 95\% CL whereas the strong mode prefers $\rm{log}(\textit{G}_{\rm{eff}}{\rm MeV}^2)=-1.36^{+0.24}_{-0.30} $ and has $\sum m_\nu =0.30^{+0.26}_{-0.25}$eV, $N_{\rm eff}=3.80^{+0.78}_{-1.0}$ at 95\% CL. The strong mode also has a higher Hubble constant, a smaller comoving sound horizon at baryon drag epoch, $r_{\rm drag}$, and a lower amplitude of the matter power spectrum, $\sigma_8$. These parameter differences compensate for the introduction of the non-zero $\textit{G}_{\rm eff}$. The strong mode gives a better consistency between {\it Planck} and {\it WMAP} in their best fitting $H_0$ and $r_{\rm drag}$ posteriors which is desirable.

It is important to consider how much the choice of prior impacts the parameters. The posterior for the strongly self-interacting neutrinos is enhanced if we impose a linear prior on \geff, as the density of points probed is higher around the region where \geff\ is non-zero. For our logarithmic prior, the parameter volume of an interacting scenario is relatively smaller. For our analysis, we chose a logarithmic prior as it does not make an explicit choice for the energy scale of the problem \cite{FCY}. 

\begin{figure}[t!]
\includegraphics[width = 0.95\linewidth]{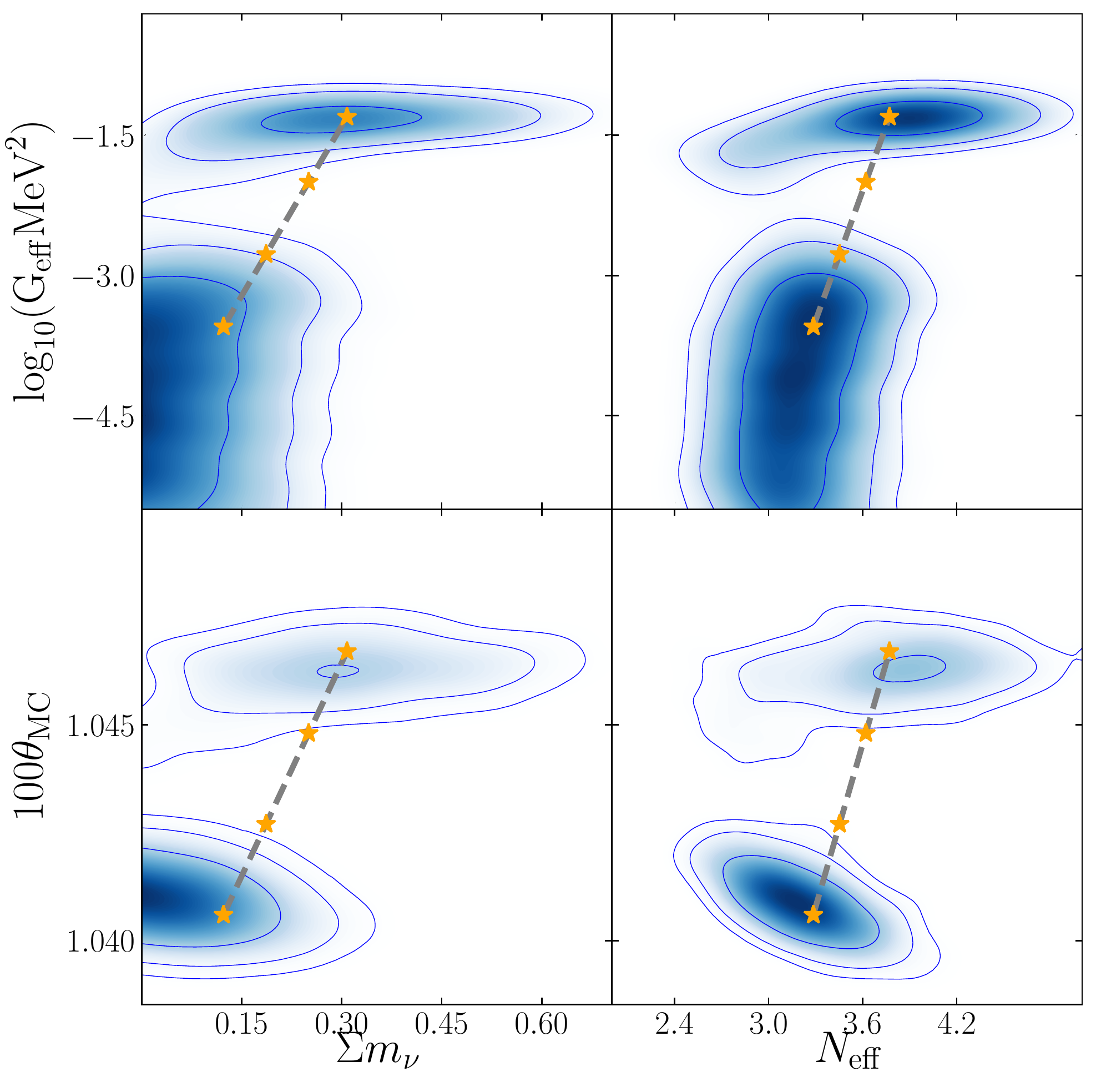}
\caption{Illustration of the line connecting the best-fitting models in each mode that we use to compute spectra and likelihoods. The orange stars are the locations of the 4 points in parameter space sampled for Figure \ref{fig:spectra} and Figure \ref{fig:difflmax}}
\label{fig:path}
\end{figure}

When using the {\it WMAP} data, which measures angular multipoles $\ell\leq 1200$, we find in Figure \ref{fig:planckparams} that the distribution is not bimodal. Instead, the neutrino self-interaction strength is consistent with zero and has an upper limit of $\rm{log}(\textit{G}_{\rm{eff}}{\rm MeV}^2) < -1.85$ at 95\% confidence. It is only when using smaller-scale data, with $\ell > 1000$, that the bimodality appears. Indeed, this bimodality was first found when combining {\it WMAP} data with data from the ACT and SPT small-scale CMB experiments \citep{FCY}. Figure \ref{fig:planckparams} also shows that the {\it WMAP} data do not favor the strongly interacting mode, implying that the smaller scale data in the $1200 < \ell \leq 2500$ range enhance the preference for the strong mode.

\subsection{CMB spectra as a function of increasing \geff}

To understand why the two models fit both datasets well, and why the central region with \geff~$\approx10^{-2.5}$~MeV$^2$ is excluded by the {\it Planck} data, we identify best-fitting models in each of the two peaks of the distribution: one with no, or low, self-interaction (essentially \LCDM), and the other with high self-interaction strength. Sampling evenly spaced points along the straight line connecting the peaks in the nine-dimensional parameter space, as shown in Figure \ref{fig:path}, we compute the likelihood of each of the datasets, and generate the CMB power spectra corresponding to each point.

\begin{figure}
\includegraphics[width = 0.95\linewidth]{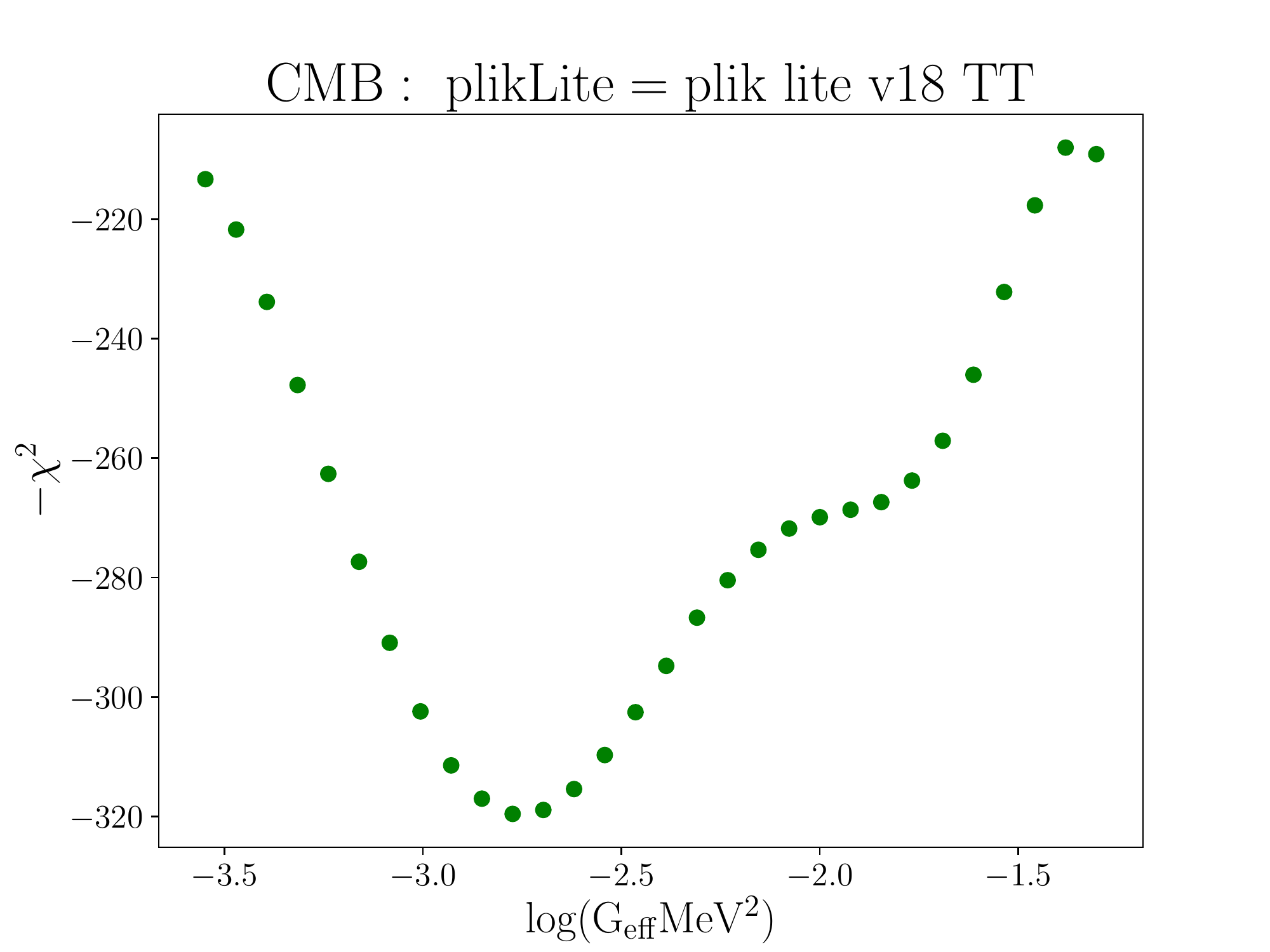}
\caption{$-\chi^2$ values for the {\it Planck} $\ell>30$ data along the path shown in Figure \ref{fig:path} to show the clear bimodality and the likelihoods between the two modes. The two modes have the same likelihood; the difference in posterior distribution for \geff\ is then due to the volume of well-fitting models in our chosen parameterization.}
\label{fig:chisq}
\end{figure}

Figure \ref{fig:chisq} shows the {\it Planck} `Plik-lite' $-\chi^2$ along this path. We find that the two modes are each similarly well fitted to the data, with $\chi^2_{\rm strong}-\chi^2_{\rm weak} \approx 6$ but there exists a valley of bad fitting in between them. This is at $\rm{log}(\textit{G}_{\rm{eff}}{\rm MeV}^2)\approx -2.75$, at which $-\chi^2$ is about 100 lower than at the two peaks.
There is a curved path between the two peaks that has a more modest reduction in likelihood: the point at which the two modes have most overlap is displaced from the line that directly connects the peaks in 9-dimensional space. We also find that the low-$\ell$ CMB temperature, the CMB lensing and the BAO $-\chi^2$ are roughly constant along the path shown in Figure \ref{fig:path}. It is the high-$\ell$ CMB data that exclude the central region and create the bimodality. 

\begin{figure*}[t!]
\includegraphics[width = 0.48\linewidth]{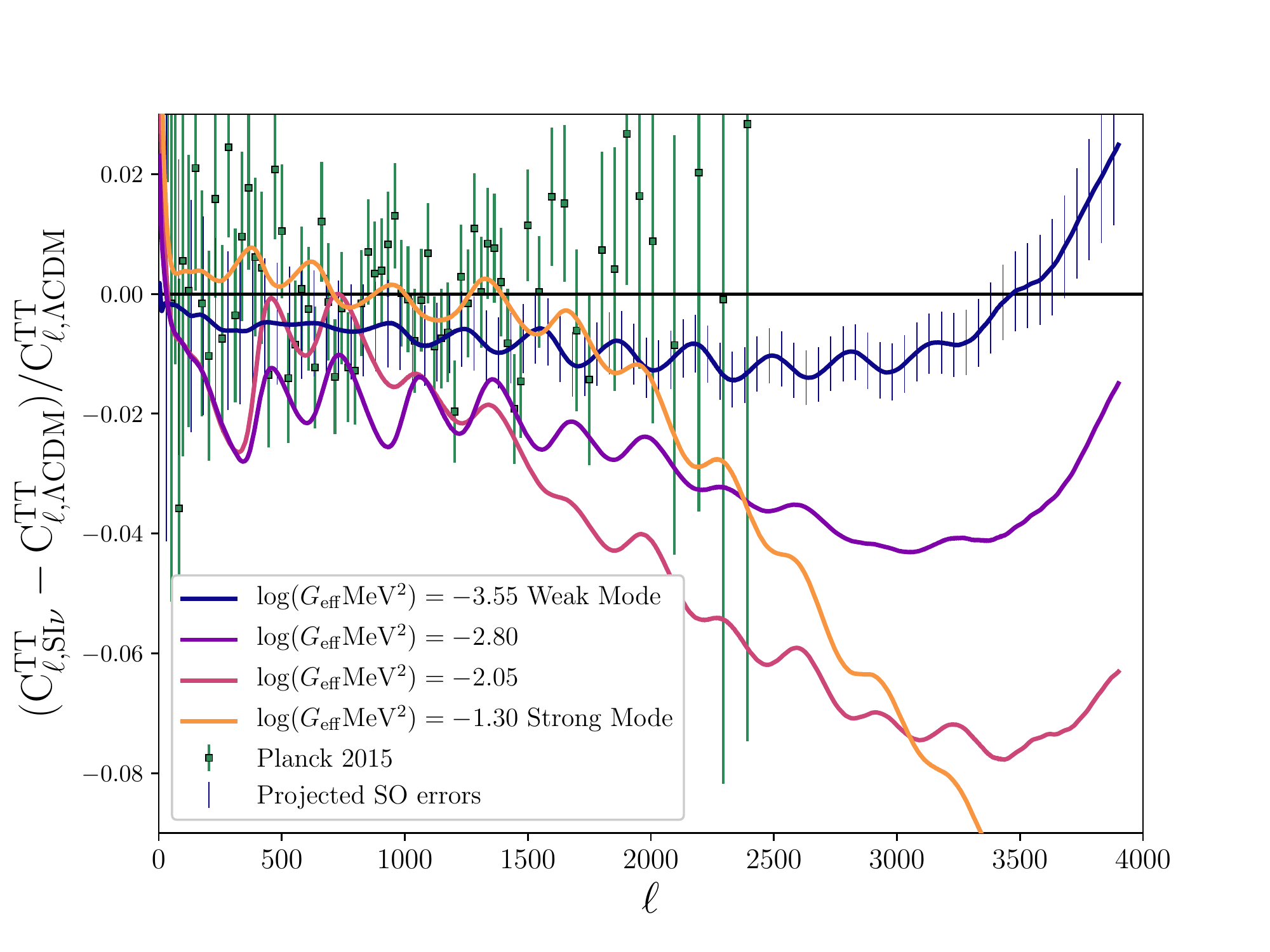}
\includegraphics[width = 0.48\linewidth]{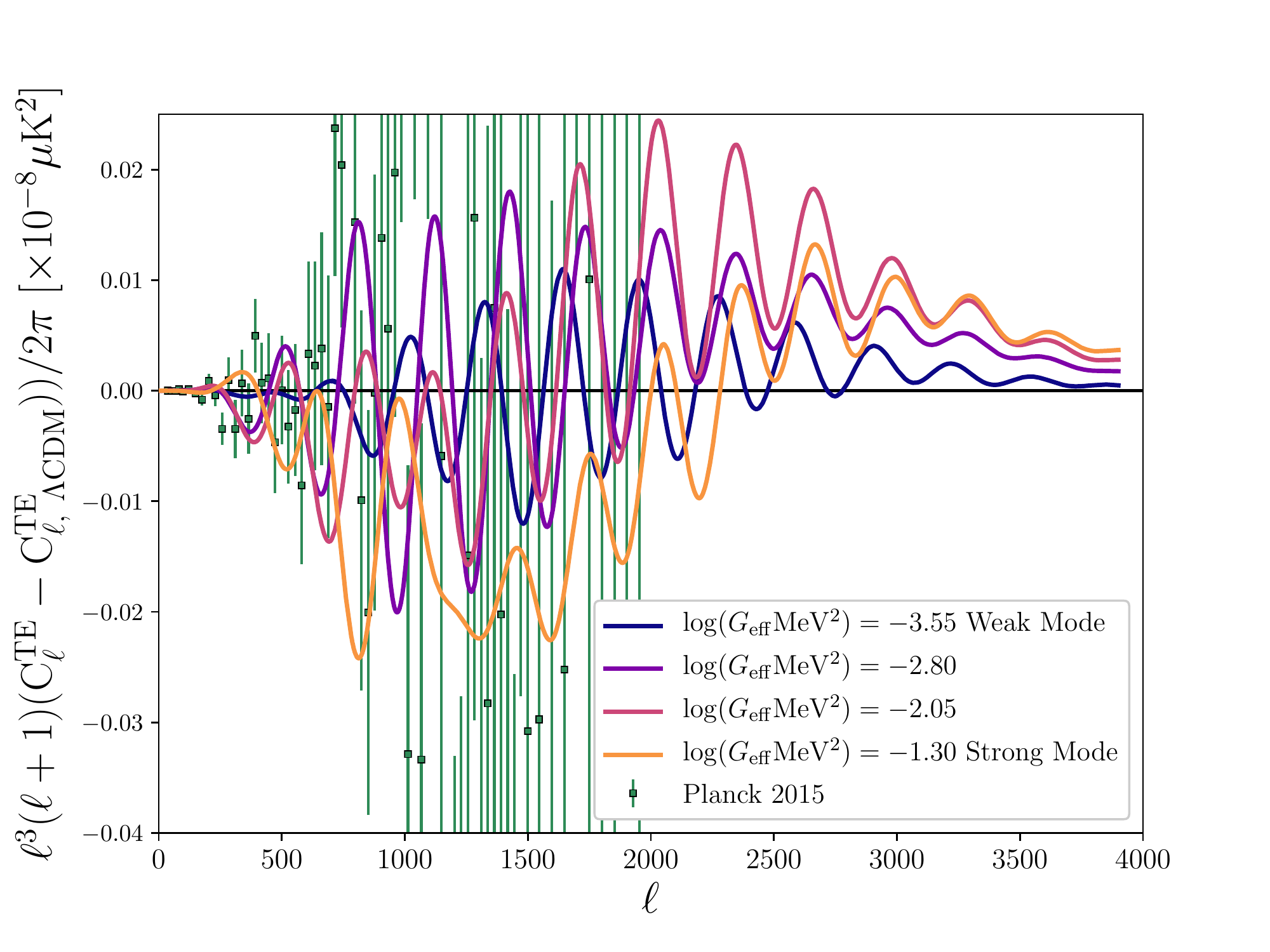}
\caption{CMB power spectra (for TT, left, and TE, right) at the points shown in Fig. \ref{fig:path}, shown as residuals compared to the best-fitting \LCDM\ model from \cite{planckparams:2018}. The {\it Planck} error bars are shown, and forecasted SO errors are indicated on the left-hand plot. The `weak' and `strong' modes both fit $\ell<2000$ data but diverge at smaller scales and differ in TE. The intermediate values for \geff\ have a lower TT power at $\ell>1000$, so are excluded by {\it Planck} data.}
\label{fig:spectra}
\end{figure*}

\begin{figure}[b!]
\includegraphics[width = \linewidth]{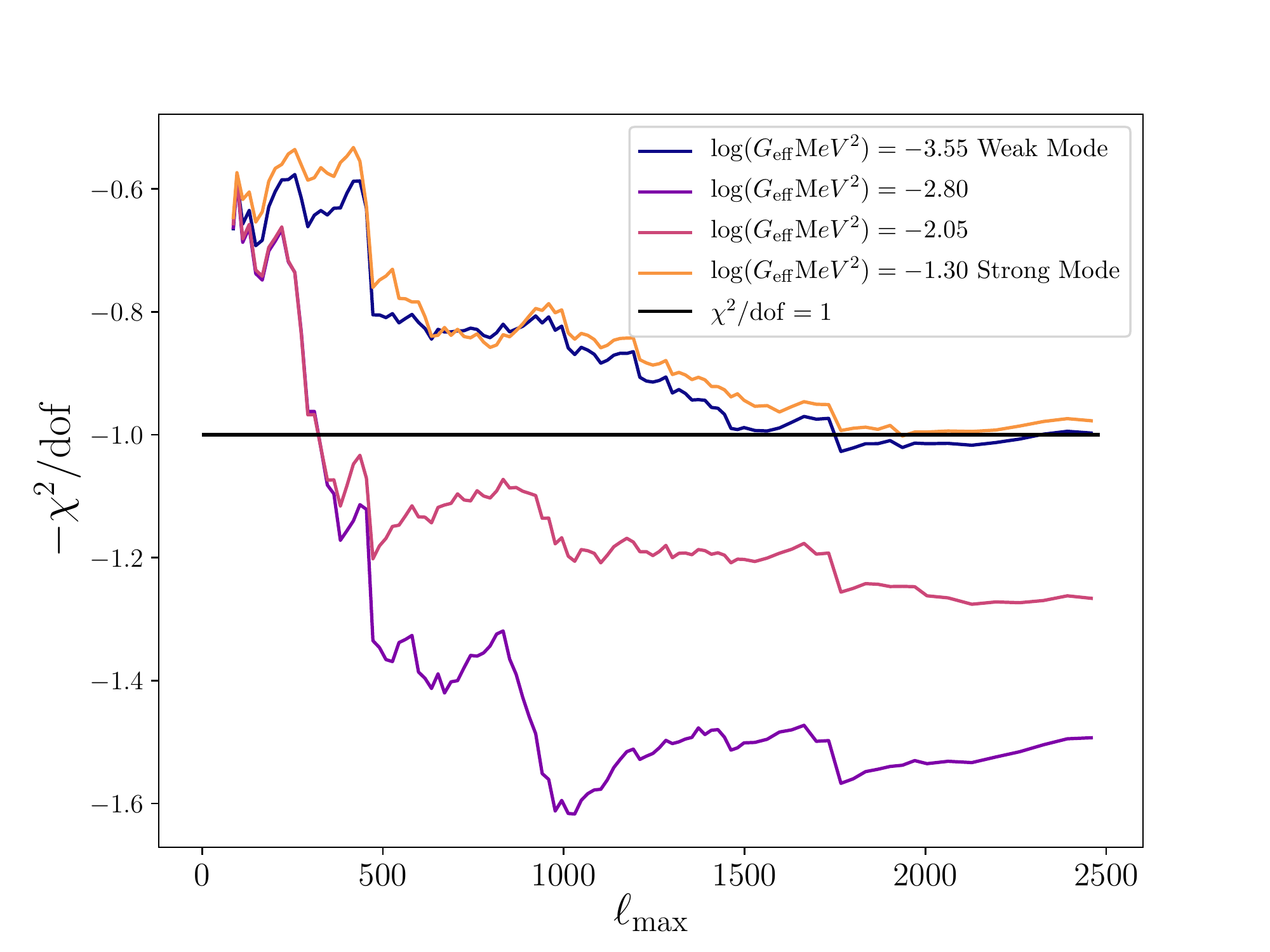}
\caption{The goodness of fit ($-\chi^2/{\rm dof}$ for $\ell>30$ from the `plik-lite' {\it Planck} likelihood) as a function of $\ell_{\mathrm{max}}$ for the models shown in Fig. \ref{fig:spectra}. This shows how models between the two well-fitting modes are poor fits to the {\it Planck} data at small scales.} 
\label{fig:difflmax}
\end{figure}

\begin{figure}[b!]
\vspace{-15pt}
\includegraphics[width =\linewidth]{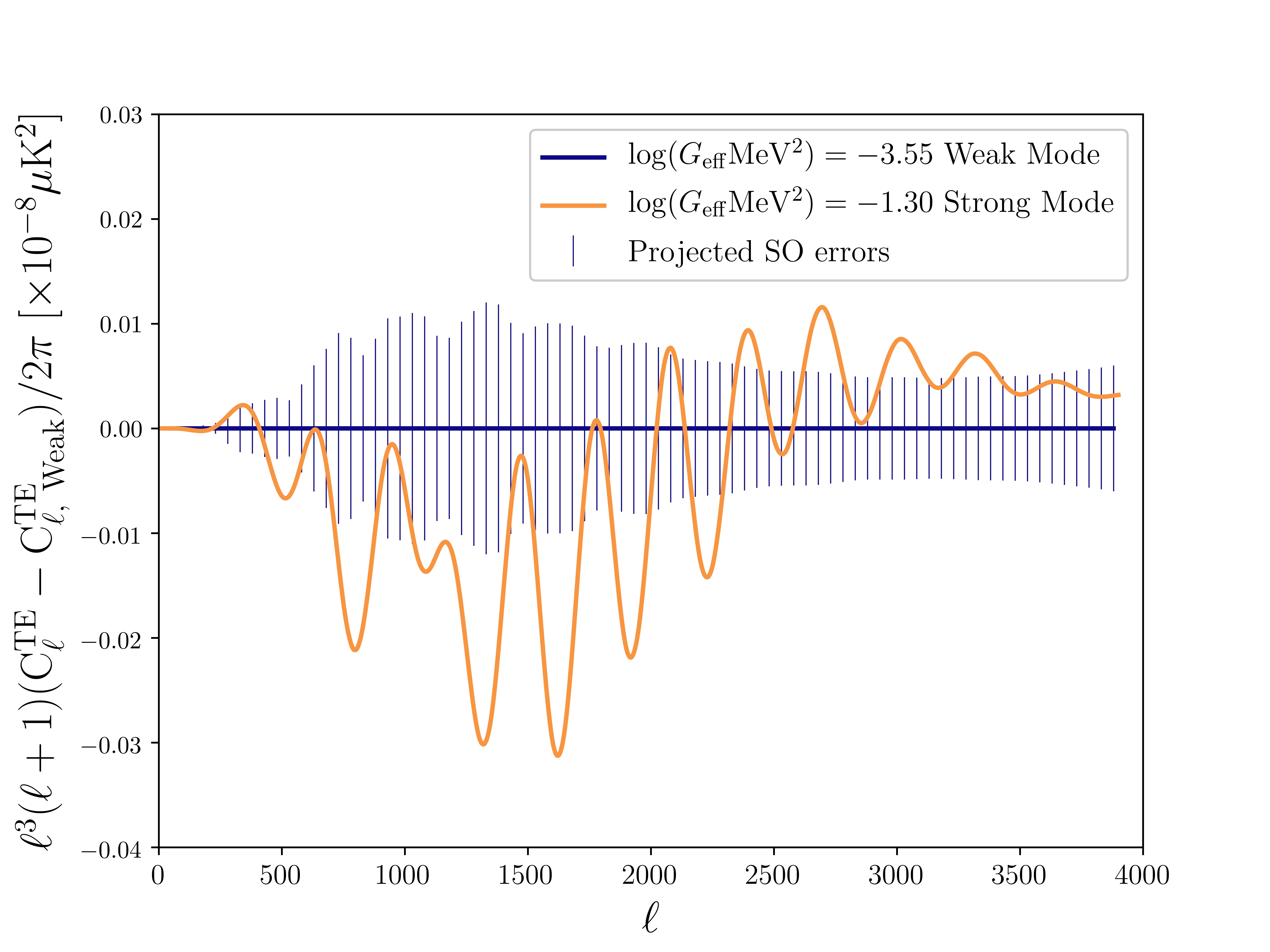}
\caption{The residual TE power spectrum between the strong and the weak best-fitting mode, together with the Simons Observatory projected errors. These data should allow the two models to be distinguished.}
\label{fig:TE}
\end{figure}
In Figure \ref{fig:spectra} we show the TT, and TE spectra for these four models with increasing \geff, showing the fractional residual between the spectra and the best-fitting \LCDM\  model.  Since {\it Planck} provides a good measurement up to $\ell\approx 2000$, the two modes fit the data well and do not show significant residuals in TT or TE.  In constrast, the power spectra corresponding to points in parameter space between the two modes that a reasonable fit to {\it WMAP} data as seen in Figure \ref{fig:planckparams} do not fit the {\it Planck} TT data at $\ell>1000$, $\ell$ range similar to that of {\it WMAP}.

We examine this scale dependence in more detail by calculating how the $-\chi^2/{\rm dof}$, $-\chi^2$ per degree of freedom
(dof), depends on the smallest scale included, $\ell_{\mathrm{max}}$, as shown in Figure \ref{fig:difflmax}. There are significant drops for the two intermediary points around $\ell_\mathrm{max}\approx$ 500 and 1000. With these two drops, the $-\chi^2/{\rm dof}$ of those two points are at $\approx-1.2,\ -1.5$ respectively, excluding them from viability. Meanwhile, the two peaks steadily approach $-\chi^2/{\rm dof}\approx-1$. This is an additional illustration that the $\ell>1000$ data prefer the two modes but disfavor intermediate interaction strengths.

\begin{figure*}[t!]
\vspace{-10pt}
\includegraphics[width = \linewidth]{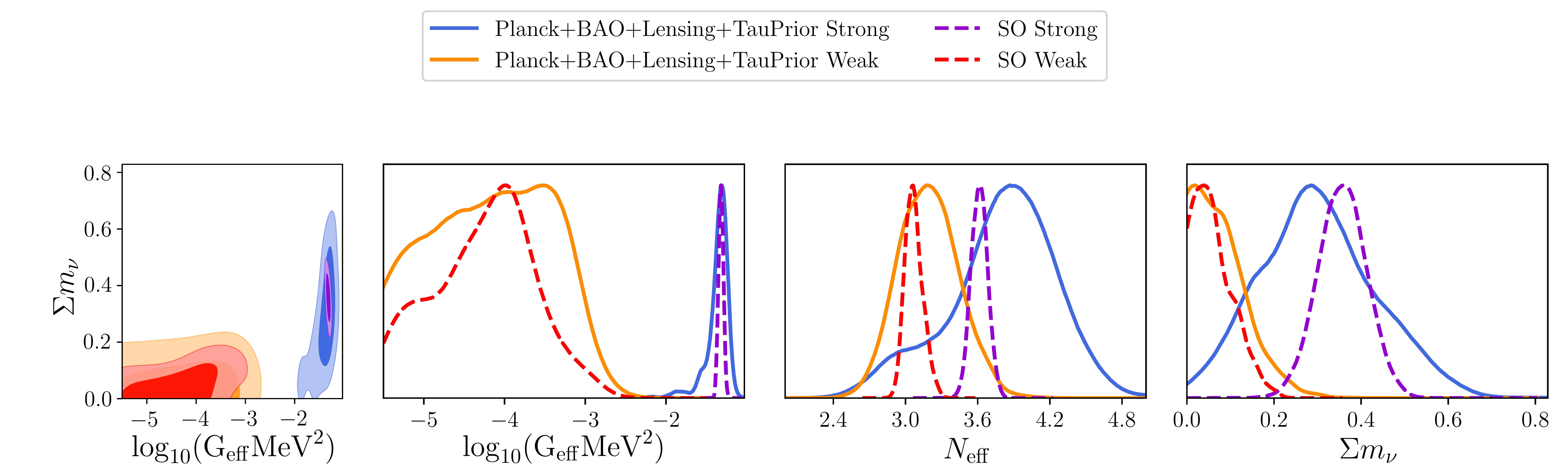}
\caption{Forecasted parameter constraints using expected Simons Observatory power spectrum measurements. We forecast for the two modes independently by searching in the parameter space assuming a cosmology described by one of the modes, then swapping the assumed cosmology for the other mode. The distributions are expected to tighten considerably compared to current {\it Planck} data, and should allow the models to be distinguished.}
\vspace{-10pt}
\label{figure:so}
\end{figure*}

\subsection{Impact of upcoming CMB data}
\label{subsec:forecast}

We explore how upcoming data from the Simons Observatory (SO) will effect the bimodality. Since the posterior is bimodal, we generate two different simulated models \footnote{Simulated data are produced using the \texttt{makeperfectforecast.py} code within \texttt{CosmoMC}.}. One simulated power spectrum is generated with the best-fitting weak mode power spectra, the other with the best-fitting strong mode power spectra, shown in Figure \ref{fig:spectra}. 

Figure \ref{fig:spectra} shows that at smaller scales than measured by {\it Planck}, the two modes diverge from each other. Including the forecasted SO uncertainties, we find that the strong mode differs from the weak mode by more than 1-sigma at $\ell>2000$. This suggests that with improved small-scale data, we could potentially rule out one of the modes. Furthermore, the two modes also have significantly different TE spectra. We show forecasted SO uncertainties on the weak mode in Figure \ref{fig:TE}, finding that the strong mode differs from the weak mode by more than 1 $\sigma$ at most minima. This indicates that the improved TE spectrum's sensitivity to the phase of the CMB power spectrum will help put further constraints on the bimodality.

The forecasted distributions for SO are shown in Figure \ref{figure:so}. The distributions are narrowed significantly compared to those shown in Figure \ref{fig:planckparams} for {\it Planck}, suggesting that SO data will be capable of further constraining the parameters for each mode. Indeed, the forecasted constraints on neutrino physics in the weak mode tighten compared to the distributions in Figure \ref{fig:planckparams} such that the upper bound of $\mathrm{log_{10}(\textit{G}_{eff}MeV^2)}$ decreases, the upper bound of $\sum m_\nu$ decreases by $20\%$, and the errors on $N_{\rm eff}$ decreases by about 70\%. For the strong mode, the errors decrease by about 80\% for $\mathrm{log_{10}(\textit{G}_{eff}MeV^2)}$, by about 60\% for $\mathrm{\sum m_\nu}$, and by 85\% for $N_{\rm eff}$.

In each case, for either the strong or weak mode as the input model, the Multinest MCMC routine does not find the other mode. In fact, the other mode is excluded by many standard deviations. So, if SO finds the data to be significantly closer to one mode than the other, the data would exclude the non-favored mode.

This is expected as SO will better measure the high $\ell$ TT data, and provide an improved measurement of the TE power spectrum data as shown in Figures \ref{fig:spectra} and \ref{fig:TE}, which are the areas of CMB power spectra data where the two models are non-degenerate. If the true model is \LCDM, the forecasted uncertainties are small enough to be able to rule out the strong mode. In contrast, if the true model is the best-fitting strongly interacting mode, SO could potentially rule out $\Lambda$CDM.

\section{Conclusion}
By comparing the probability distributions in the parameter space using {\it WMAP} and {\it Planck} data we show that the {\it Planck} data in the angular range $1000<\ell<2500$ allow a model with strongly interacting neutrinos, and disfavor a model with more moderate interactions. We explore this in more detail by looking at the power spectra and at the likelihood of the data for increasing \geff.
We highlight that high $\ell$ TT, and improved TE data will be pivotal in constraining or ruling out the bimodality. The Simons Observatory will make these measurements and is forecasted to significantly improve constraints. If the data were to favor it, SO would be capable of ruling out the bimodality. The strong mode has cosmological parameters that are significantly different to \LCDM, including a higher Hubble constant, lower amplitude of structure, higher neutrino mass and higher effective neutrino species. While the particular model considered here is ad-hoc, further exploration of physical models for the neutrino sector seem warranted.

\acknowledgements
We thank David Spergel and Lyman Page for useful comments and Olivier Dor\'e for early comments. M. P. acknowledges the support of the Department of Astrophysical Sciences at Princeton University. C. D. K. acknowledges support from the National Science Foundation award number DGE1656466 at Princeton University. F.-Y. C.-R.~acknowledges the support of the National Aeronautical and Space Administration (NASA) ATP grant NNX16AI12G at Harvard University. Part of this research was carried out during the Undergraduate Summer Research Program at the Department of Astrophysical Sciences, Princeton University. This is not an official SO Collaboration paper.

\bibliography{nu_ref}  
\end{document}